\documentclass{article}

\usepackage{amsmath}
\usepackage{amssymb}
\usepackage{graphicx}

\usepackage[super,comma,sort&compress]{natbib}

\begin{document}
\title{Local electron heating in nanoscale conductors}

\author{Roberto D'Agosta\footnote{Email: dagosta@physics.ucsd.edu} , Na Sai and Massimiliano Di Ventra\footnote{Corresponding author. Email: diventra@physics.ucsd.edu}\\
Department of Physics, University of California - San
Diego, La Jolla, CA 92093}

\maketitle

\begin{abstract}
The electron current density in nanoscale junctions is typically
several orders of magnitude larger than the corresponding one in bulk
electrodes. Consequently, the electron-electron scattering rate
increases substantially in the junction. This leads to local
electron heating of the underlying Fermi sea in analogy
to the local ionic heating that is due to the increased electron-phonon
scattering rates. We predict the bias dependence of local electron
heating in quasi-ballistic nanoscale conductors,
its effect on ionic heating, and
discuss possible experimental tests of our results.
\end{abstract}

In the process of electrical charge transport, the dissipation of
energy via heat production plays a significant role. This effect is
of particular importance in nanoscale systems, like e.g., atomic or
molecular structures between bulk electrodes,\cite{DiVentra2004}
since it determines their structural stability under current flow.
It is now understood that the large current densities in
nanojunctions, compared to their bulk counterparts, may lead to
substantial heating of the nanostructure ions.
\cite{Todorov1998,Chen2003,Huang2006} This effect is directly
related to the consequent increase of the electron-phonon scattering
rates in the junction. In quasi-ballistic systems, i.e. when the
mean free path is much longer than the dimensions of the
nanostructure, by assuming a bulk lattice heat conduction mechanism,
the local ionic temperature is predicted to be $T_{ion}\propto
\sqrt{V}$, where $V$ is the applied
bias.\cite{Todorov1998,Chen2003,Huang2006} For the same reasons,
and due to the viscous nature of the electron
liquid,\cite{Conti1999} we here suggest that the local increase of
the electron-electron scattering rate in the junction gives rise to
local heating of the {\em underlying Fermi sea}, whether the system
has one or many conducting channels. This local electronic
temperature would also affect the electron-phonon scattering rates,
and consequently the bias dependence of the ionic temperature.

In this Letter, we first estimate the bias dependence of the local
electron temperature in quasi-ballistic systems, assuming no ionic
heating is produced. We will then determine the effect of the electron
heating on the bias dependence of the ionic temperature. We finally
discuss possible experiments to test our predictions.

Let us start from a simple argument on the expected bias dependence
of the electron heating. Assume first no electron-phonon scattering
is present in the system. The power generated in the nanostructure
due to exchange of energy between the current-carrying electrons and
the underlying Fermi sea has to be a small fraction of the total
power of the circuit ${V^{2}}/{R}$ ($R$ is the resistance). Let us
define this fraction as $P=\alpha{V^{2}}/{R}$, with $\alpha$ a
positive constant to be determined from a microscopic theory. At
steady state this power has to balance the thermal current $I_{th}$
(heat per unit time) carried away into the bulk electrodes by the
electrons. Let us first assume that the electron thermal
conductivity $k$ follows a bulk law, i.e. it is related to the
specific heat per unit volume $c_V$ via the relation $k= v_F
\lambda_e c_V/d$ where $v_F$ is the Fermi velocity, $\lambda_e$ is
the electron mean free path, and $d$ the dimensionality of the
system.\cite{prec1}
We will later give a general argument to justify
the form of this law in quasi-ballistic systems away from electronic
resonances. We know that at small temperatures the specific heat of
an electron liquid is proportional to the electronic temperature
$T_{e}$ so that
\begin{equation}
k=\gamma T_e
\label{kbulk}
\end{equation}
where $\gamma=k_F^2 k_B^2 \lambda_{e}/9 \hbar$ in 3D, and $\gamma=\pi k_{F}k_{B}^{2}\lambda_{e}/6\hbar$ in 2D, $k_F$ is the Fermi
momentum, and $k_B$ the Boltzmann constant . The thermal current, $I_{th}\propto k T_{e}$, is then given by
$I_{th}=\gamma' T^2_e$ where $\gamma'$ has to
be determined from a microscopic theory. At steady state the condition $P=I_{th}$ implies a linear increase of the electronic temperature with bias,
\begin{equation}
T_e=\gamma_{e-e}V, \label{Testimate}
\end{equation}
if the coefficient $\gamma_{e-e}$ does not vary appreciably with
bias. In macroscopic electrical contacts this result is known as the
``$\varphi-\Theta$ relation".\cite{Holm1967} 
Here, we will re-derive it from a
microscopic theory and determine the quantity $\gamma_{e-e}$ in
terms of the properties of the junction and of the electron liquid.
This theory needs to take into account the influence of
electron-electron interactions on both the heat production and
dissipation. Two of the authors (RD'A and MDV) have recently shown
that the dynamics of the electron liquid in nanojunctions can be
described using a hydrodynamic approach,\cite{DAgosta2006a}  (see Ref.~(9) for a different 
hydrodynamical approach to the electron liquid flow) so that
the equations of motion for the electron density $n(r,t)$ and
velocity field $v(r,t)$ can be written in the Navier-Stokes form
\cite{Landau6,DAgosta2006a}
\begin{equation}
\begin{split}
D_{t}n(r,t)=&-n(r,t)\nabla \cdot v(r,t)\\
mn(r,t)D_{t}v_{i}(r,t)=&-\nabla_{i}P(r,t)+\nabla_{j}\sigma_{ij}(r,t)\\
&-n(r,t)\nabla_{i}V_{ext}(r,t)
\label{ns}.
\end{split}
\end{equation}
where $P$ is the pressure, $D_{t}=\partial_{t}+v\cdot\nabla$ is the
convective derivative operator, and $V_{ext}$ is an external potential
(like the electron-ion potential). The stress tensor $\sigma_{i,j}$
is ~\cite{DAgosta2006a,Vignale1997b,Conti1999,Tokatly2005a}
\begin{equation}
\sigma_{i,j}=\eta\left(\partial_{i}v_{j}(r,t)+\partial_{j}v_{i}(r,t)-\frac{2}{d}
\delta_{ij} \nabla\cdot v\right)+\zeta\delta_{ij} \nabla\cdot v,
\label{sigmaxc}
\end{equation}
where $\eta$ and $\zeta$ are positive parameters which correspond to
the shear and bulk viscosity of the liquid, respectively. They have
been determined for the electron liquid using linear response
theory.\cite{Conti1999} Since in the DC limit (the regime of
interest here) $\zeta \ll \eta$,\cite{Conti1999} in what follows we
will consider the shear viscosity only. This term is the one
responsible for the heating of the Fermi sea: electrons crossing the
junction experience an internal ``friction'' with the electrons of
the underlying Fermi gas, thus creating electron-hole pairs and
generating heat. In keeping with this hydrodynamic picture we
supplement Eqs.~\ref{ns} for the particle densities and velocity
field with an equation which describes the energy dissipation and
diffusion. Since the quantity $\nabla_i \sigma_{i,j}$ gives the
force acting on the electron liquid caused by viscosity, the
equation for the heat transfer reads~\cite{Landau6}
\begin{equation}
\begin{split}
T_{e}(r,t)D_{t}s(r,t)=&\sigma_{i,j}(r,t)\partial_j
v_i(r,t)\\
&+\nabla\cdot [k(r,t) \nabla T_e(r,t)]
\end{split}
\label{heat}
\end{equation}
where $k$ is again the thermal conductivity and $s$ is the local
entropy per unit volume.\cite{prec2}

For metallic quantum point contacts (QPC) of interest here the electron liquid can be assumed incompressible. \cite{Sai2005} At steady state eq.~\ref{heat} therefore becomes
\begin{equation}
\sigma_{i,j}(r)\partial_j
v_i(r)+\nabla\cdot [k(r) \nabla T_e(r)]=c_{V}(T_{e})v(r)\cdot \nabla T_e(r).
\label{heatequilibrium}
\end{equation}
where we used $T(r)\nabla s(r)=c_{P}\nabla T(r)$, $c_{p}$ being the
specific heat per unit volume at constant pressure.\cite{prec3} For an electron liquid at low temperatures $c_{V}\sim
c_{P}$.\cite{prec4} Equations \ref{ns}
and~\ref{heatequilibrium}, with all quantities time-independent,
together with the definition of the stress tensor \ref{sigmaxc},
constitute a set of equations that describes the charge and heat
flow at steady state. In this effective theory quantum mechanics
enters explicitly in the electron liquid constants $\eta$, $k$, and
$c_{V}$.\cite{prec5}

Before proceeding with the calculation of the temperature from
eq.~\ref{heatequilibrium} we need to know the form of the thermal
conductivity $k$. We give here a simple argument to justify the use of
the bulk form, eq.~\ref{kbulk} in quasi-ballistic systems at low
temperatures (and bias) and in the absence of electronic
resonances. Following the Landauer approach,\cite{Mellokumar} let us
assume the electrons tunnel across the junction from the
right electrode which is in local equilibrium with temperature $T_R$
and chemical potential $\mu_R$, to the left electrode in local equilibrium  with temperature $T_L$ and chemical potential $\mu_L$.
If the transmission
coefficient for electrons to move elastically from right to left is ${\cal T}(E)$, the energy current transported by the
charge is of the form
\begin{equation}
 I_{th}\propto \int dE~E {\cal T}(E)[f_R(E,T_R)-f_L(E,T_L)]
\label{heatT}
\end{equation}
where $f_{R(L)}$ is the local Fermi-Dirac distribution function of
the right (left) electrode at its own local temperature and chemical
potential. Let us set $T_L=0$ and consider a small right electrode
temperature $T_R \equiv T_e$. For small biases, in the absence of
electronic resonances, the transmission coefficient ${\cal T}(E)$
can be assumed independent of energy, ${\cal T}(E)\equiv {\cal T}$.
From eq.~\ref{heatT} we therefore see that $I_{th}\propto T_e^2$,
i.e. the thermal conductivity must be of the form~\ref{kbulk}. In
the following we will assume ${\cal T} \sim 1$, typical of metallic
QPCs, and therefore assume the form~\ref{kbulk} for $k$ with the
same coefficients.
\begin{figure}
\includegraphics[width=7cm,clip,angle=-90]{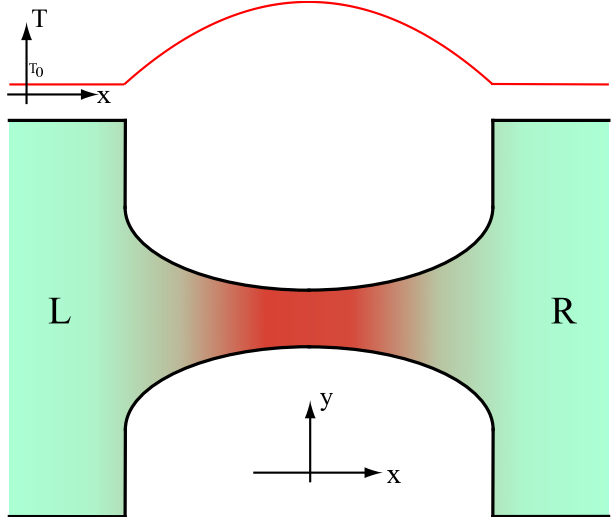}
\caption{Schematic of the local electronic temperature profile of a
nanojunction.}
\label{fig1}
\end{figure}

In order to obtain the temperature profile of the electron gas
inside the nanojunction and evaluate the relation between $T_e$ and
$V$, we need to solve eq. \ref{heatequilibrium} together with
Navier-Stokes equations~\ref{ns} with the boundary conditions
imposed by the geometry of the junction. For an arbitrary system,
this is obviously an impossible task. Since we are only interested
in the analytical dependence of the temperature on bias, and the
estimate of the coefficients in this relation, we proceed as
follows. Let us assume a given profile for the velocity of the fluid
and solve eq. \ref{heatequilibrium} with the boundary condition
that the temperature reaches the constant bulk value far away from
the constriction. We choose a reference frame in which the direction
of the current flow is along the $x$ axis and the constriction,
whose length is $L$, is centered at $x=0$ (see Fig.~\ref{fig1}). Let
us also assume that the velocity varies negligibly in the $(y,~z)$
directions compared to the much larger variation along the $x$
direction.

Since the system has a variable cross-section and the total current
$I$ has to be constant throughout the device, the fluid velocity and
the cross-section of the system are approximately related via
\begin{equation}
I=en v_{x}(x)A(x).
\label{current}
\end{equation}
The cross-section profile $A(x)$ contains the
information about the geometry of the device. In the following we
assume the simplest profile possible which allows an analytical solution: the adiabatic model (see Fig.~\ref{fig1})
\begin{equation}
A(x)=
\left(A_c+r x^{2}\right)\theta\left(\frac{L^2}{4}-x^2\right)+ 
\left(A_c+r\frac{L^{2}}{4}\right)\theta\left(x^2-\frac{L^2}{4}\right)
\label{profile}
\end{equation}
with $A_c$ and $r$ positive parameters describing the minimal cross-section of the junction and the
``rate'' at which it opens, respectively, and $\theta(x)$ is the Heaviside step function.\cite{prec6}
With this choice of the profile $A(x)$ we get
the velocity profile we need to supply in
eq. \ref{heatequilibrium} to obtain the temperature profile
\begin{equation}
v_{x}(x)=\frac{I}{ne}\left[
\frac{1}{A_c+rx^{2}} \theta\left(\frac{L^2}{4}-x^2\right)+
\frac{1}{A_c+rL^{2}/4}\theta\left(x^2-\frac{L^2}{4}\right)\right]
\label{velocity}.
\end{equation}

The total electron velocity $u(x)$ is the sum of the Fermi velocity and
the fluid velocity, $u=v_{F}+v_{x}$.
In the bulk where the heating can be considered
negligible, $u$ is related to the applied bias voltage
via $m u^{2}=2(E_{F}+eV)$, and since in general,
$v_{x}\ll v_{F}$ we get $v_{x}=eV/m v_{F}$ for the fluid velocity in the bulk.\cite{prec7}
This is the boundary condition that eq. \ref{velocity} has to satisfy for $|x|\ge L/2$, i.e.
$A_c+rL^{2}/4=mv_{F} I/n e^{2}V$. This condition implies $rL^{2}/4A_c\sim 1$ which simply reflects the physical
fact that, due to screening, the electron velocity approaches the bulk value very fast away from the junction.

In the following we assume that deep inside the electrodes the
electron gas is in equilibrium with a zero-temperature thermal bath,
i.e. $\lim_{|x|\to\infty}T_{e}(x)=0$.\cite{prec8} The solution to
eq.~\ref{heatequilibrium} can be calculated analytically and
expressed in terms of a combination of rational, trigonometric and
hypergeometric functions. From this solution we can estimate that
the contribution of the the advection term $c_{V}v\cdot\nabla T$ is
controlled by the dimensionless quantity
$\Gamma\equiv(Id/v_{F}\lambda_{e}n e A_c)\sqrt{A_c/r}$. Due to the
quasi-ballistic assumption, we find $\Gamma \ll 1$ so that the
advection term can be neglected.

In the absence of the advection term the solution to
eq.~\ref{heatequilibrium}
inside the constriction ($|x|<L/2$)
is given by
\begin{eqnarray}
T_e^2(x)&=& \left(\frac{I}{neA_c}\right)^2
\left[\frac{(d-1)\eta}{3 d \gamma}\right]
\left\{
\frac{1-rL^2/4A_c}{(1+rL^2/4A_c)^2}-\frac{1-rx^2/A_c}{(1+rx^2/A_c)^2}\right.\nonumber\\
&&\left.+3\sqrt{\frac{r}{A_c}}\left[\frac{L}{2}\tan^{-1}\left(\sqrt{\frac{r}{A_c}}\frac{L}{2}\right)
-x \tan^{-1}\left(\sqrt{\frac{r}{A_c}}x\right)\right]\right\}.
\label{tprofile1}
\end{eqnarray}
Let us then estimate the maximum temperature which, in the absence of the advection term, occurs at $x=0$,
\begin{equation}
T_M=
\left(\frac{I}{neA_c}\right)\sqrt{\frac{d-1}{3d}\frac{\eta}{\gamma}}
\sqrt{
3\sqrt{\frac{r}{A_c}}
\frac{L}{2}\tan^{-1}\left(\sqrt{\frac{r}{A_c}}\frac{L}{2}\right)
+\frac{1-rL^2/4A_c}{(1+rL^2/4A_c)^2}-1
}~.
\label{tbar}
\end{equation}
In the linear transport regime and for ${\cal T}\sim 1$,  $I=G_0 V$
with $G_{0}=2e^{2}/h$. As we have anticipated in
eq.~\ref{Testimate}, the relation between the temperature and the
bias then simplifies to $T_M=\gamma_{e-e} V$, where $\gamma_{e-e}$
is a constant that can be read from eq.~\ref{tbar}. It is
interesting to note that a similar expression has been found in the
case of superconducting junctions, where, however, the maximum
temperature inside the junction is found to be independent of both
the junction geometry and the electron
properties.\cite{Tinkham1977}

By using the expression of $\eta$ as a function
of the particle density given in Ref.~(5)
we can estimate the value of the constant $\gamma_{e-e}$ for various systems.
We consider here both a 3D gold point contact and a 2D electron gas (2DEG).
For a 3D gold ($r_s=3$) QPC with effective cross-section $A_c=7.0$ \AA$^{2}$, by assuming a typical inelastic mean free path in
quasi-ballistic systems of $\lambda_{e}\sim 100~\mathrm{nm}$,  we get
$\gamma_{e-e}(\textrm{QPC})\simeq 65~ \mathrm{K}/\mathrm{V}$. For a 2DEG, assuming $r_s=10$, $\lambda_{e}\simeq 10~\mathrm{\mu m}$, and  $A_c=20~\mathrm{nm}$ we get
$\gamma_{e-e}(2DEG)\simeq 1.2\times10^{2} ~\mathrm{K/V}$.

Let us now discuss the effect of local electron heating on the ionic
heating. In the 2DEG we expect a negligible heating of the ions. In
atomic-scale junctions, instead, the ions may heat up substantially
due to electron-phonon scattering. As discussed at the beginning of
the paper, the ionic temperature in quasi-ballistic systems is
predicted to be $T_{ion}=\gamma_{e-p} \sqrt{V}$, if no electron
heating is taken into account. The constant $\gamma_{e-p}$ has been
estimated in \cite{Todorov1998,Chen2003} for atomic and molecular
systems. If we now allow for the electron heating, then part of the
ionic heating is lost in favor of the local electron temperature.
Since the source of energy is the same for both processes (the
kinetic energy of the electron), the ionic temperature must be {\em
lower} in the presence of the electron heating than without it. To
first order we can assume the electron-electron and electron-phonon
processes independent and occurring with equal probability. Let us
also suppose that, for a given bias, the local electron temperature
in the absence of electron-phonon scattering, $T_e$, is smaller than
the ionic temperature in the absence of electron heating, $T_{ion}$.
The energy lost by an electron due to heating is ``seen'' by the
phonons as a ``sink'' of energy at the electron temperature $T_e$.
This energy however is ``lost'' by the phonons at their own rate.
The balance between the power absorbed by the phonons in the
junction and the power dissipated away at steady state therefore
gives the new ionic temperature
\begin{equation}
T_{w}=\left[\gamma_{e-p}^{4} V^{2}- \gamma_{e-e}^{4}V^{4}\right]^{1/4}.
\label{Tcorr}
\end{equation}
If $T_{ion}\simeq T_{e}$ the processes
of energy release from current-carrying electrons to the phonon gas or to the Fermi sea cannot be considered independent and a microscopic theory that considers them on equal footing is needed.

We point out that for metallic QPCs the theoretical values for
$\gamma_{e-p}$ are of the order of hundreds of
$\mathrm{K}/\mathrm{V}^{1/2}$  while our
estimate for $\gamma_{e-e}$ is of the order of tens of
$\mathrm{K}/\mathrm{V}$.\cite{Todorov1998,Chen2003} We thus expect that, for these systems,
the electron heating contribution to the ionic temperature is small
for a wide range of biases. In the case of a 2DEG, on the other
hand, the ionic heating is either not present or it is very small
and it may be possible to measure the effects of the electron
temperature directly. In the case of organic molecules between
metallic electrodes it is known that $\gamma_{e-p}$ decreases
exponentially with the length of the molecules.\cite{Chen2003} For
such systems, we thus expect the two effects to be of similar
importance and one should therefore be able to observe deviations
from the $\sqrt{V}$ dependence of the ionic temperature due to
electron heating as predicted in eq.~\ref{Tcorr}. However, the
exact value of the parameter $\gamma_{e-e}$ requires generalization
of the present theory to the case of non-ideal conductance, i.e.,
for a transmission coefficient ${\cal T}$ less than 1. This can, in
principle, be done but no general analytical solution can be
derived. The reason is that one needs to know both the thermal
conductivity dependence on the transmission coefficient, and the
junction profile that leads to such coefficient. For ${\cal T}\ll 1$
this may even lead to a non-linear dependence of $T_{e}$ as a
function of $V$. We can however comment that if one works at fixed
current while varying ${\cal T}$, for example, by acting with a gate
voltage or by stretching the chemical bonds of the molecule, an
increase of electron temperature due to reduced thermal conductivity
is expected. We finally suggest that the local increase of the
electron temperature in a nanojunction may be extracted by measuring
the Johnson-Nyquist noise.\cite{Djukic2006} Any deviation of this
noise compared to the expected one in the absence of electron
heating may give a direct evidence of this effect.

In conclusion, we have discussed the phenomenon of local electron heating in nanoscale junctions. We
have predicted its bias dependence and estimated, using a hydrodynamic approach, its magnitude in
quasi-ballistic systems. We have also discussed its role on the local ionic heating and suggested systems
where this effect is most likely to be observed.

{\bf Acknowledgments.} We gratefully acknowledge useful discussions with G. Vignale. This work has been supported by the
Department of Energy grant DE-FG02-05ER46204.


\begin{thebibliography}{99}
\bibitem{prec2}Equation~\ref{heat} can also be derived from first principles for the electron liquid applying a similar approach that was used to derive the quantum Navier-Stokes equations~\ref{ns}
[see Ref.~(11)].
\bibitem{prec3}In
eq.~\ref{current} we are assuming that the temperature variations
are small so that the density fluctuations due to temperature can be
neglected.
\bibitem{prec4}We assume that the system is in local thermal
equilibrium, thus local thermodynamic quantities such as energy and
entropy densities can be defined.
\bibitem{prec7}In this
hydrodynamic picture, the Fermi velocity does not contribute to the fluid velocity, the
former being an ``incoherent'' part of the electron
motion.
\bibitem{prec8}
If the electron temperature inside the electrodes, $T_0$, is not zero, the 
electron temperature is $T^2(x) = T^2_0 + T^2_e (x)$, with $T^2_e(x)$ given in eq. 
11.
\end{thebibliography}
\end{document}